\documentclass{article}
\usepackage{fortschritte}
\def\beq{\begin{equation}}                     % 
\def\eeq{\end{equation}}                       %
\def\bea{\begin{eqnarray}}                     %         %
\def\eea{\end{eqnarray}}                       %       % 
\begin {document}                 
\def\titleline{Quasi  Normal Modes}
\def\email_speaker{Ivo Sachs
{\tt ivo@theorie.physik.uni-muenchen.de}}
\def\authors{Ivo Sachs}
\def\addresses{Theoretische Physik,
Ludwig-Maximilians Universit\"{a}t\\ 
Theresienstrasse 37\\
D-80333, M\"{u}nchen\\
Germany}
\def\abstracttext{In this talk we review different applications of quasinormal modes to 
black hole physics, string theory and thermalisation in quantum field theory. In particular, we describe the the relation between quasi normal modes of AdS black holes and the time scale for thermalisation in strongly coupled, large $N$ conformal field theory. We also discuss the problem of unitarity  within this approach. }
\large
\makefront
\section{Introduction}
It is well established that black 
holes behave like thermodynamical systems. In particular, at equilibrium, the various thermodynamical quantities, such as temperature and entropy, 
are determined in terms of mass, charge and angular momentum of the black  
hole. However, a direct derivation of these properties from first 
principles requires a detailed knowledge of the microscopic theory of black 
holes which is still poorly understood. 

This is in contrast to finite temperature field theory which is a well 
studied and conceptually well defined problem (e.g. \cite{Fetter}). 
For a small perturbation, the process of thermalisation is described by linear 
response theory \cite{Fetter}. The relaxation process is then 
completely determined by the poles, in the momentum space representation,  
of the retarded correlation function of the perturbation.

It turns out that these seemingly different phenomena in black hole physics 
and finite temperature quantum field theory may, in fact well be related 
to each other at a fundamental level. The idea that quantum field theory 
might provide a microscopic definition of quantum gravity is not new and 
has been explored, for instance in induced gravity and composite gravity 
models. However, the most concrete realisation of this idea is found in 
string theory which realises a very specific correspondence 
between gravity in anti- de Sitter space and quantum field theory in flat space, \cite{Maldacena:1997re,Gubser:1998bc,Witten:1998qj}  (see \cite{Aharony:1999ti} for a review). In this correspondence 
certain field theory correlators can be equivalently described by solving 
Green function equations in a dual gravitational background. This suggests that 
gravity may be interpreted as a {\it masterfield} for certain correlation 
functions in quantum field theory. 

If we accept that this 
correspondence between gravity and quantum field theory also holds at finite 
temperature, this provides an effective tool to relate thermal properties of black holes and finite temperature quantum field theory.  Although, in general, neither the field theory correlators, nor the Green functions in the corresponding gravity background are known explicitly, non-trivial information about thermalisation in quantum field theory can nevertheless be be obtained by studying small perturbations of a black hole away form equilibrium. Such perturbations are described by 
quasinormal modes \cite{Frolov:wf}. 
These oscillations are similar to normal modes of a closed system. 
However, since the perturbation can fall into the black hole or radiate to 
infinity the corresponding frequencies are complex.  In the linear regime a given perturbation can be expressed in terms of a free field of mass $m$ and  spin $s$ propagating in the black hole background. For instance in asymptotically flat space, a scalar perturbation $\Phi$ of mass $m$ satisfies the equation of motion
\begin{equation}
\label{QN1}
\left(\nabla^2- m^2\right)\Phi=0\\,
\end{equation}
with ingoing and outgoing boundary conditions at the horizon and infinity respectively,
\begin{equation}
\Phi=\cases{\Phi^{in}&$r=r_S$\cr \Phi^{out}&$r=\infty$\\.}
\end{equation}
In general there are no closed solutions to (\ref{QN1}) but for the general discussion in this section the qualitative behaviour of the solutions will be sufficient. For asymptotically flat four dimensional geometries the time dependence of $\Phi$ is of the form 
\begin{equation}
\Phi(t)\simeq\cases{e^{-i\omega t}&small $t$\cr
t^{-\alpha}& large $t$\\,}
\end{equation}
where $\alpha=2l+3$ depends on the angular momentum. 
The power law tail describes the scattering off the asymptotic $1/r$ potential while the 
exponential describes the 'quasinormal ringing'. The quasinormal frequencies $\omega=\omega_R+i\omega_I$ have a negative imaginary part proportional to the temperature of the black hole. In particular, in asymptotically flat space-time $\omega_I$ is independent of the perturbation thus encoding information about the black hole only. 

For black holes in anti-de Sitter (AdS) space time the situation is different. The cosmological constant leads to a growing potential in 
the asymptotic region effectively acting as a confining box. On the other 
hand, perturbations can still fall into the black hole and correspondingly 
the oscillations are still damped. The corresponding frequencies can be 
determined by a variety of analytical and numerical methods\footnote{See \cite{SN} for further references} 
\cite{Horowitz:1999jd,Birmingham:2001pj,SN,QNM}. Via the gravity/quantum field theory correspondence described above quasinormal modes 
should have an interpretation in the corresponding quantum field theory. 
It has been suggested in \cite{Horowitz:1999jd} that the quasinormal frequencies 
provide the time scale for thermalisation of perturbations in the 
corresponding finite temperature field theory. As explained above, 
in quantum field theory this process is described by linear response 
theory.  The precise connection has now been established for 
the case of anti-de Sitter black holes and 
will be described in section 2.  In section 3 we then discuss the relation of quasinormal modes with the problem of unitarity.  

\section{Quasi normal modes and linear response theory}
With the formulation of the AdS/CFT between (super) gravity on $(d+1)$-dimensional 
anti-de Sitter space and conformal field theory in $d$ dimensions the prospect of a quantitative interpretation of quasinormal modes  of black holes in asymptotically anti-de Sitter space became a reality. The boundary conditions for the quasinormal wave equation have to be such that the perturbation cannot escape to infinity. In particular, the power law tail is absent in anti-de Sitter space. 
Another modification with respect to the asymptotically flat space is the existence of an extra scale, the curvature radius $\ell$ of AdS space. However, for very massive black holes, $r_S>\!\!>\ell$, it is not hard to see \cite{Horowitz:1999jd} that the quasinormal frequencies are proportional to the temperature, $T_H$, of the black hole. The first few quasi normal frequencies where computed numerically in \cite{Horowitz:1999jd}  for scalar perturbations  of AdS-Schwarzschild black holes in four-, five- and seven dimensions (see also \cite{SN}) for further results and references). According to the AdS/CFT correspondence, the imaginary part of $\omega$ should have an interpretation as the time scale for the  thermalisation of small perturbations in the dual conformal field theory. In quantum field theory this process is described using linear response theory \cite{Fetter}: For a perturbation $H_{pert}=\int J(x){\cal{O}}(x)$, the deviation of the expectation value $\langle{\cal{O}}\rangle$ is then expressed in terms of the retarded correlation function $G_R(t-t',{\underline{x}}-{\underline{x}}')$ as  
\begin{equation}
\delta\langle{\cal{O}}\rangle=\int dt' d^{d-1}x'\, G_R(t-t',{\underline{x}}-{\underline{x}}')J(t',{\underline{x}}')\, .
\end{equation}
For simplicity we consider an instantaneous and homogenous perturbation $ J(x,t) = \delta(t) $. One then finds 
\begin{equation}
\delta \langle {\cal O}(x,t)\rangle = \int_{-\infty}^{\infty}
\frac{d\omega}{2 \pi} e^{-i\omega t}\tilde G_{R}(\omega, 0)=-i\sum\limits_{ \omega_i: poles}\hbox{Res}(\tilde G_{R})|_{\omega_i} e^{-i\omega_i t}\ .
\end{equation}
Thus the quasinormal modes should correspond to the poles of the retarded Green function in the dual conformal field theory. That this should be so can also be seen directly from the AdS/CFT correspondence. Indeed, Keski-Vakkuri\footnote{See also \cite{Son:2002sd}.} \cite{Danielsson:1999zt} argued that in Fourier space the retarded CFT two-point correlator for the operator ${\cal{O}}$ dual to a scalar field $\Phi$ propagating in AdS is given by 
\begin{equation}
\tilde G_{R}(\omega,0) =\frac{\Phi^-(\omega)}{\Phi^+(\omega)}\ ,
\end{equation}
with boundary conditions such that $\Phi$ is ingoing at the horizon and $\Phi^\pm$ are the asymptotically divergent and vanishing part of $\Phi$ at infinity respectively. With this definition it is then clear that the poles of $\tilde G_{R}$ are given by the points $\omega$  for which $\Phi^+$ vanishes. This, in turn is just the quasinormal mode boundary conditions\footnote{The precise boundary condition is that the current has to vanish at infinity \cite{Birmingham:2001pj,SN}} and thus the quasinormal frequencies are just the poles of the retarded  correlation function of the dual operator. This correspondence can be tested explicitly in AdS$_3$ since in this case the both, the quasi normal modes as well as the poles of the dual correlator can be calculated and compared explicitly. Since in the dual $2$-dimensional conformal field theory the left and right moving degrees of freedom decouple we expect that each sector has its own sequence of quasinormal modes. This is indeed the case and the two sets of quasinormal frequencies are given by 
\begin{eqnarray} 
%\label{PFD1} 
\omega_L&=&k-4\pi i T_L(n+h_L)  \ ,\nonumber\\ 
\omega_R&=&-k-4\pi i T_R(n+h_R)\ . 
\label{10} 
\end{eqnarray} 
Here $n=0,1,2,...$ takes integer values, $T_{L/R}$ are the temperatures of the left and right moving sectors  related to the Hawking temperature via $\frac{1}{T_L}+\frac{1}{T_R}=\frac{1}{T_H}$. Finally, $h_{L/R}$ are the conformal weights of the 
perturbation operators in the conformal field theory. These are related to the mass and spin of the perturbation of the AdS$_3$-black hole by the usual relation \cite{Aharony:1999ti} 
$h_R+h_L=\Delta$, $h_R-h_L=\pm s$ with $\Delta=1+\sqrt{1+m^2}$.  
This  set of poles characterises the decay 
of the perturbation on the CFT side, 
and coincides precisely with the quasinormal frequencies 
of the BTZ black hole for perturbations of various mass and spin \cite{Birmingham:2001pj}. 

In dimensions bigger than two the conformal Ward-identities are not enough to determine the thermal $2$-point correlators exactly. However, the quasinormal  modes can still be determined numerically thus allowing to extract non perturbative information about the resonances in strongly coupled, large $N$  conformal field theory at finite temperature. In particular, N\'u\~nez and Starinets \cite{SN} found that the quasinormal frequencies  for scalar perturbations of AdS$_5$-Schwarzschild black holes can be written in the form\footnote{This also holds for massless vector perturbations  with $\Delta=\Delta_-$ and for massless gravitons \cite{SN}.} 
\begin{equation}\label{s}
\omega\simeq2\pi T\left(n+\frac{\Delta}{2}-\frac{1}{2}\right)(1-i)\ ,\qquad n=0,1,2,...
\end{equation}
which the provides a prediction for the poles of the retarded correlators of the dual operators in strongly coupled conformal field theory. It is interesting to compare  this set of poles with the retarded $2$-point function of a free massless scalar field  ($\Delta =1$) in four dimensions. The poles of this correlator are located at $\omega= 2\pi i T n$ which agrees with the imaginary part in (\ref{s}). In particular the shift by $-\frac{1}{2}$ in (\ref{s}) is required by the periodicity of the finite temperature correlation function. The temperature dependent real part of the quasinormal frequencies on the other hand is a novel effect which should be interpreted as a 
feature of strongly coupled conformal field theory.

\section{Finite Size Effects and Unitarity}
So far we have assumed that the conformal field theory, dual to the AdS-black hole background lives in infinite volume. However, if the volume is finite we expect the thermalisation process to be qualitatively different on the field theory side. The question is then how this can be matched in the gravitational description. For concreteness let us consider the  BTZ black hole which is locally  AdS$_3$ with curvature radius $\ell$.  According to the AdS/CFT correspondence this background is dual to the world volume theory of a system of $k=\frac{\ell}{8 G_N}$ parallel $D1$ and $D5$ branes. This theory can be interpreted as a gas of
strings that wind around a circle of length $L$ with target space
$T^4$.
The individual strings can be simply or multiply wound such that the
total
winding number is $k$. In the large $k$ limit this theory undergoes a first order phase transition \cite{Maldacena:1998bw} at $\beta_c=L$. The low temperature phase consists of $k$ simply would strings while the high temperature phase is described by a single string wound $k$ times.  This transition is mimicked on the gravity side by the Hawking-Page transition \cite{Hawking:1982dh} between AdS-black hole at large temperature and thermal AdS at low temperature. Since the effective volume of the field theory is finite in the low temperature phase the energy spectrum is discrete and consequently the real time two point correlation function 
has poles on the real line in momentum space (e.g. \cite{Fetter}) corresponding to a (quasi) periodic function in real time. This is just a manifestation of the well known Poincar\'e recurrences which are in turn a consequence of the finite volume in phase space. On the gravity side, the poles on the real line are in $1-1$ correspondence with {\it normal} modes in thermal AdS \cite{Birmingham:2002ph}
\begin{eqnarray}
\label{set}
\omega = \pm\frac{4 \pi}{L}(n+h)\ .
\label{poles}
\end{eqnarray}
The Hawking-Page transition is thus a transition between
oscillatory behaviour (normal  modes) at low temperature  and  exponentially decaying
behaviour (quasinormal modes) at high temperature.

The question is then what happens a finite $k$? One possibility is to approximate the full retarded correlator by the sum of the two dominant contributions at large $k$ \cite{Maldacena:2001kr}
\begin{equation}
\langle{\cal{O}}(w,\bar{w}) {\cal{O}} (w',\bar{w}')\rangle
=
e^{-S_{\rm BTZ}}\langle{\cal{O}} ~{\cal{O}}'\rangle_{\rm BTZ}+
e^{-S_{\rm AdS}}\langle{\cal{O}} ~{\cal{O}}'\rangle_{\rm AdS}\ , 
\end{equation}
up to a global normalisation. This implies a lower bound on the correlation which is required by unitarity. However, from the field theory point of view the volume of the field theory at finite $k$ is always finite since even multiply wrapped strings have finite length. One thus expects an oscillatory (quasi-periodic) behaviour for any temperature. This in turn is not compatible with the existence quasinormal modes which clearly imply an exponential decay of the retarded correlator. 
We can go some way to resolve this puzzle by recalling that at finite $k$ there cannot be a  phase transition in any two dimensional conformal field theory. Consequently the field theory will always be in the low temperature phase no matter what temperature. On the CFT side the free energy can be determined exactly for any $k$ \cite{Dijk} and one sees explicitly how the discontinuity arises at $k=\infty$. In the classical gravity approximation on the other hand one finds a Hawking-Page transition for any $k$. Thus, if the the AdS/CFT correspondence holds for any $k$, then the string corrections to the metric must remove the horizon. Ways to implement such corrections have been analysed for instance in\footnote{See also E. Rabnovic's talk at this conference.} \cite{Lunin:2001jy,Barbon:2003aq}.

\section{Discussion}
We have seen that quasinormal modes provide an effective and quantitative  tool to explore thermalisation in large $N$ conformal field theory. In particular, the imaginary part of the quasinormal frequencies correspond to the poles of the retarded 
thermal correlation function of various field theory observables. On the other hand, for finite the strict $N$ quasinormal modes are in conflict with unitarity on the field theory side suggesting that black boles should be interpreted as as duals of strongly coupled CFT's only in the strict $N\to \infty$ limit. An interesting question is whether the large $N$ approximation can still be used to obtain some quantitative information about the finite $N$ system for instance along the lines explored in \cite{Sundborg:1999ue,Aharony:2003sx}. 

Recently it has been suggested that the real part of the quasinormal frequencies 
may also have a quantitative interpretation in terms of the micro states of the quantised black hole, at least in asymptotically flat space. Concretely, for Schwarzschild black holes the asymptotic form of real part of $\omega_n$ approaches $T_H\ln(3)$, for $n\to\infty$. Using Bohr's correspondence principle, Hod \cite{Hod:2000it} suggested that energy levels of the quantised black hole should be given by $\Delta E=\frac{\hbar}{8\pi M}\ln(3)$ which mean that the area should be quantised as $\Delta A=4\ell_p^2\ln(3)$. Comparing this quantitative relation with the loop quantum gravity prediction fixes the otherwise undetermined Immirzi parameter \cite{Dreyer:2002vy}.

%%%%%%%%%%%%%%%%%%%%%%%%%%%%%%%%%%%%%%%%%%% 
{\bf Acknowledgement} I would like to thank D. Birmingham and S. Solodukhin for collaboration on some of the work reviewed in this talk as well as S. Theisen and V. Suneeta for helpful discussions. I would also like to extend my thanks to the organisers of the Ahrenshoop meeting for creating a productive and enjoyable environment. 
%%%%%%%%%%%%%%%%%%%%%%

\end{document}